\documentclass[conference, letterpaper]{IEEEtran}
\ifCLASSINFOpdf
\else
\fi
\ifCLASSINFOpdf
   \usepackage[pdftex]{graphicx}
\else
\fi

\usepackage[cmex10]{amsmath}
\usepackage{color}
\usepackage{fancyhdr}
\usepackage[caption=false,font=footnotesize]{subfig}

\renewcommand{\thispagestyle}[2]{}

\fancypagestyle{plain}{
        \fancyhead{}
        \fancyhead[C]{first page center header}
        \fancyfoot{}
        \fancyfoot[C]{first page center footer}
}
\begin{document}

%
\title{A Centralized Reputation Management Scheme for Isolating Malicious Controller(s) in Distributed Software-Defined Networks}

\author{\IEEEauthorblockN{Bilal Karim Mughal}
\IEEEauthorblockA{Department of Computer Science\\
Bahria University\\
Karachi, Pakistan\\
Email: bilalkmughal@gmail.com}
\and
\IEEEauthorblockN{Sufian Hameed}
\IEEEauthorblockA{Department of Computer Science\\
National University of Computer and Emerging Sciences\\
Karachi, Pakistan\\
Email: sufian.hameed@nu.edu.pk}
\and
\IEEEauthorblockN{Ghulam Muhammad Shaikh}
\IEEEauthorblockA{Department of Computer Science\\
Bahria University\\
Karachi, Pakistan\\
Email: gm.shaikh1@gmail.com}}

\maketitle

\begin{abstract}
Software-Defined Networks have seen an increasing in their deployment because they offer better network manageability compared to traditional networks. Despite their immense success and popularity, various security issues in SDN remain open problems for research. Particularly, the problem of securing the controllers in distributed environment is still short of any solutions. This paper proposes a scheme to identify any rogue/malicious controller(s) in a distributed environment. Our scheme is based on trust and reputation system which is centrally managed. As such, our scheme identifies any controllers acting maliciously by comparing the state of installed flows/policies with policies that should be installed. Controllers rate each other on this basis and report the results to a central entity, which reports it to the network administrator.
\end{abstract}
\providecommand{\keywords}[1]{\textbf{\textit{Index terms---}} #1}

\IEEEpeerreviewmaketitle

\section{Introduction}
Software-defined networks (SDN) separate the data plane and control plane from each other, contrary to traditional networks in which both are embedded in the same hardware piece [1]. In SDN, a network administrator has to implement a policy only in the controller, which is then replicated across the network's forwarding devices [2]. Despite immense popularity and ever increasing growth in deployment of SDNs, much research is needed as to whether SDN can be deployed on a large scale, that too with all the security [3]. 

Various researchers including [4, 5] acknowledge control layer as a highly vulnerable section of the SDN, which, if compromised, can result in losing entire network to a malicious entity. However, the models proposed for deploying SDNs so far do not answer one basic question: how to identify malicious or rogue controllers within a network, and how to prevent them from causing damage [6].

In this paper, we propose a scheme to enhance controller security in a multi-controller environment. Our framework identifies malicious/rogue controllers by finding out if a mismatch exists between the flows which should be installed in the switches by the controllers and those which are actually installed. We make use of a centralized trust and reputation scheme inspired by [7], in which controllers are rated positive or negative by other controllers according to their performance. The results are then reported to a central entity called the Trust Collector which aggregates the results and passes them on to the network administrator. Earliest detection of rogue controllers through such reputation management will ensure the isolation of rogue controllers before they can damage the network.

For enabling controllers to rate each other, we modified the Ryu controller code [8]. We also introduced two novel components, Policy Distributor and Trust Collector for managing trust and reputation, and for providing a benchmark to controllers against which they can compare the installed flows. The scheme was implemented in Emulab Network Emulation Testbed [9]. Initial evaluations show that our scheme is successful in identifying rogue controllers.
 
The rest of the paper is organized as follows. In Section 2, we briefly review the security threats to SDN controllers, and state-of-the-art. Section 3 discusses our architecture, components and the scheme flow. Section 4 talks about the implementation and evaluation. Section 5 sheds light on the need of introducing the central components. Section 6 concludes this paper after discussing the direction of our future work.

\section{Literature Review}

In SDN, the controllers are an easy target and are open to exploitation through unauthorized access. If the controller platform is not secure, an active adversary can hijack the network by deceiving the network devices. DNS servers are prone to these kind of attacks, shown by [10]. An entire network can be brought down if an adversary gains control of the network by hijacking the controller in this way [11]. 

Several threat vectors exist when it comes to security of the control plane. These include attacks on control plane communications, i.e. controller-controller or controller-switch communications. Apart from this, one should also be vigilant about certain higher-level applications which have access to network information through controller APIs because such applications can reprogram a network without causing much of a suspicion [11]. A major challenge here is to differentiate between the legit and malicious applications to allow/deny access. The authors in [5] argue that commonly used intrusion detection systems (IDS) might not prove to be completely useful in securing the controllers from misuse, as it may be difficult to ascertain which events resulted in the malicious behavior, and whether it should be labeled as malicious at all.

There are several studies which have tried to resolve controller security problems in SDN. For example, the Security-Enhanced Floodlight (SE-Floodlight) controller provides a mechanism for authentication of applications, role-based authorization for avoiding conflicts in flow-rule insertion, and conflict detection and resolution [12]. It does not, however, address one core problem, that is, isolating a compromised controller in a distributed environment. 

SDNs are logically centralized networks in which a single controller maintains multiple switches and other network devices, but in case of a man-made or technical mishap, this proves to be a single point of failure too [13]. To overcome this, distributed architectures like DISCO [14] have been proposed in which multiple controllers manage the network for better resilience and faster network management. Some network architectures such as HyperFlow [15] and Onix [16] distribute the control plane physically, but keep it logically centralized. 

The distributed systems described above, however, do not take into account the security aspects. For example, they do not provide a comprehensive framework for identifying and isolating a malicious controller out of several others. On the other hand, so-far proposed schemes for securing the control layer do not discuss the feasibility of their solutions in the distributed environments. To the best of our knowledge, no concrete work has been done to resolve this problem, and therefore this is an open challenge for research.

\section{Architecture}

The objective of our work is to develop a framework for singling out a malicious controller in distributed SDN. We achieve this by employing a trust and reputation scheme among controllers. We are working on a distributed controller environment in which the secondary controllers are deployed not as a dormant backup but as active load-balancers. However, for either use case, the controllers need to have access to all switches, so that in case one controller goes down due to an act of sabotage or for any other reason, the other controllers can prevent disruptions in network environment. 

In our architecture, controllers rate each other after verifying the policies installed by them in switches against the policies that are dictated by a central entity called the Policy Distributor. The Policy Distributor is a component introduced by us for consistent policy enforcement throughout the distributed SDN. The second component specific to our scheme is the Trust Collector, which asks controllers to rate their peer controllers and takes ratings from them. The code for both the components was written in Python and they were deployed as separate components. We describe the working of individual components below.

\subsection{Components}

\subsubsection{Policy Distributor}

It contains all of the policies that are to be installed by the controllers. Conventionally, a network administrator defines the policies directly into the controller, but in our scheme, a network administrator defines the policies in the Policy Distributor. These policies are then periodically pushed to all of the controllers in the network. This ensures network-wide consistency as there is only one place where the policies need to be defined, thereby centralizing the administration of a distributed SDN. We used a HashMap for policy assignments which takes arguments (Controller, Policy). Copy of this HashMap can be retrieved by all controllers when needed, but every controller installs it in only the switches directly under its control. 

This helps them later in verifying whether other controllers have installed correct policies or not, and is also good for fault tolerance; in case a controller goes down, other controllers will automatically know which flow rules were in effect in the affected controller. It is assumed that the Policy Distributor is secure and protected from hijacking, and any changes made to it are purely intentional.

\subsubsection{Trust Collector}

It is another central entity which is responsible for trust management. After the Policy Distributor has pushed the policies to the controllers, the Trust Collector after a specific duration, asks all the controllers of the network for their opinion about their peer controllers. Specifically, it asks other controllers to check whether their peer controllers have installed the policies in switches as dictated by the Policy Distributor, or they have (maliciously) installed different policies. The controllers then initiate their respective Policy Checkers (discussed in next section) and fetch the flow tables from the switches. If a controller finds any discrepancy between the flow tables fetched from switches and the policies sent by the Policy Distributor, it reports the results to the Trust Collector. We use the flow tuple format to specify and compare policies, e.g. policy1: \{srcIP='8.8.8.8', action='drop'\}.

\begin{figure*}[t!]
\centering
\setlength\fboxsep{0pt}
\setlength\fboxrule{0pt}
\fbox{\includegraphics[width=7.0in]{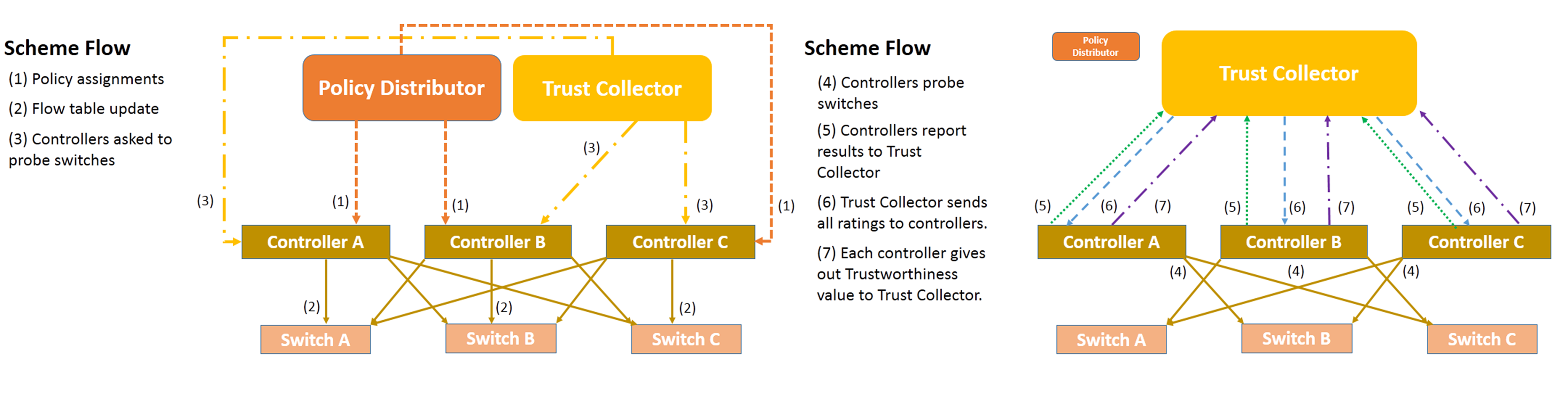}}
\caption{Flow diagram for the entire scheme. Shows message exchange flow between switches, controllers, trust collector, and policy distributor.}
\label{f7}
\end{figure*}

\subsubsection{Policy Checker}

We introduce another simple component called the Policy Checker, integrated in the Ryu controller. The primary purpose of Policy Checker is to simply probe the switches to fetch the installed policies, so that they can be compared with the policies sent out by the Policy Distributor. 

\subsection{Trust Calculation}

The mechanism of trust collection in our scheme is based on PET Model [7], however, we have made necessary changes to their method to suit our environment. The PET model is designed for strict P2P environments where there is no central entity, and the nodes are dependent on ratings obtained from each other to calculate trustworthiness. In our scheme, however, we have a central entity called the Trust Collector, which collects individually calculated trustworthiness values from all controllers and presents it to the operator for review. 

When the Trust Collector asks controllers to find out any mismatch between policies installed and policies that had to be installed, the controllers start probing the switches. At this point, all the controllers simultaneously act as recommender and recommendee. A recommender who finds out a mismatch flags the recommendee based on following function.
\[
  \ h(x)=\begin{cases}
               S_1 , x=G , S_1 > 0\\
               S_2 , x=B, S_2 < 0 \hspace{0.5cm} and \hspace{0.5cm} |S_2| > S_1
            \end{cases}
\]

Where G and B are the constants used for match and mismatch, respectively. In case of a match, a score of S1 is output, whereas in case of mismatch, S2 is given as output. The rating output by the hash function is then used in calculating the recommendation Er. Note that we use G to represent good behavior, similar to PET model [7], but we use B to represent bad behavior while PET model uses it to represent byzantine behavior.

Figure 2 shows the different parameters that go in to calculation of the trustworthiness value. The recommendation value Er for a controller A is the average value of recommendations that other controller have given to A. Therefore, in order to calculate Er for, let's say, controller A, controller B will need access to recommendations that other peers have given to A. The Trust Collector helps here by allowing all controllers to send their calculated recommendations about other controllers to itself. Once all recommendations are at the Trust Collector, each controller can then retrieve the (global) accumulation of all recommendations about any given controller from the Trust Collector.

The second thing the controllers need to calculate is the interaction-derived information Ir. In the PET model, Ir is a special recommendation given by a peer A to other peers based on how good or bad of a service those other peers have provided to only peer A, that is, unlike Er, Ir does not take into account the recommendations from other peers.

Our controller environment is slightly different from the pure P2P environment assumed by the PET model, since in our environment no controllers directly provide any services to other controllers as in a P2P system, so we changed the meaning of Ir such that Ir is now each controller's individual recommendation about its peer controllers based on whether they have installed policies in the switches correctly or not. Thus Ir is an individual controller's own opinion about a given controller A and it does not take into account what other controllers say about A. This saves Ir from getting overwhelmed if a majority of controllers (maliciously) rate controller A as negative.

The Er and Ir values are finally used to calculate the reputation Re in a weighted fashion such that,

\newcommand\tab[1][1cm]{\hspace*{#1}}

\tab W(Er) = 0.2
	
\tab W(Ir) = 0.8

The values are based on suggestions from the PET model. A higher value for W(Er) would mean that we put a lot of trust in the environment but since we consider our environment risky, therefore we have not set a very high value.

The purpose of reputation Re is to accumulate the past and current values of a controller's performance. That is, the reputation value is the historical accumulation for a recommendee's past behavior from the recommender's viewpoint. It will reflect the overall quality of the recommendee for a long time period. For example, if a controller which is being rated has installed 99 correct policies but 1 incorrect policy due to, let's say, a software bug, then that controller's reputation doesn't immediately become completely negative. Rather, the final reputation value is calculated through a combination of current and past recommendations from both individual and collective group of controllers.

\begin{figure}[h!]
  \includegraphics[width=\linewidth]{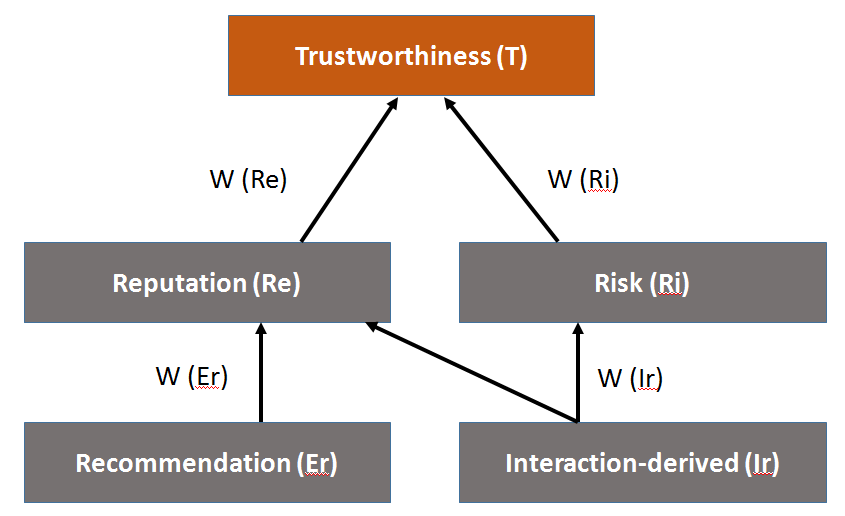}
  \caption{Trust calculation model based on [7]. Nodes collect final trustworthiness value based upon number of factors.}
  \label{fig:boat1}
\end{figure}

Since Ir gives us the personalized view of a node for its peers, therefore the PET model uses only Ir to calculate the risk value Ri for the network. This results in each controller having its own Ri value that represents its own view of the risk in the network.

The reputation Re and risk Ri values are used by the controllers to calculate trustworthiness T values. Each controller thus generates one trustworthiness value that gets collected by the Trust Collector. On PET model's suggestions, we set the weight of reputation and weight of risk to 0.5 in all controllers for calculating the T value such that: 

\tab W(Re) = 0.5
	
\tab W(Ri) = 0.5

The Trust Collector accumulates all these trust values it receives from controllers. It then averages all the trustworthiness values, and notifies the network administrator as to which controllers are malicious since their trustworthiness value was very low or which controllers are good since their trustworthiness value was high. 

\subsection{Scheme Overview}

In presence of the Policy Distributor, Trust Collector, and Policy Checker integrated within the controllers, our scheme progresses as follows: 

We have a network with three controllers and three switches, such that each controller directly administers two switches. The network is in a full mesh setting, so that all the controllers have access to all switches for backup. Assuming that the network has just booted, and the switches do not have any flow rules as of now. A network administrator defines a policy in the Policy Distributor that all traffic originating from IP address 8.8.8.8 is to be dropped.   

After some time, the Trust Collector asks controllers to probe all the switches to find out if there is a mismatch between installed policies, and those dictated by the Policy Distributor. The controllers then run their respective Policy Checkers over the network. As shown in Figure 1, each controller probes switches managed by other controllers too. 

When the probe has finished and matches/mismatches have been found, each controller gives out a ratings map for every other controller, which contains good or bad scores for them. Three controllers will generate three such maps, such that in case of three controllers A, B and C, controller A will report about B and C, controller B will do it for A and C, and controller C will do it for A and B. All of these ratings are sent to the Trust Collector.

Once the Trust Collector has received the reports from all of the controllers, it combines all of them and sends back to all of the controllers, so that A will receive reports of B and C about each other, B will receive reports of A and C, and C will receive reports of A and B. Each of the controllers now has information about what its peer controllers think about other controllers. This information helps a controller in calculating average value of recommendation (Er) for other controllers. 

Er combined with Ir are used to calculate reputation Re. Combined with risk Ri, the Re is used to calculate final trustworthiness as: 

T = Reputation (Re) * Weight of Reputation [W (Re)] + Risk (Ri) * Weight of Risk [W (Ri)]

Each controller outputs trustworthiness values for other controllers. The results are fed to the Trust Collector, which aggregates the results from all the controllers and shows it to the network administrator for review. 

\section{Evaluations}

We built a prototype implementation in Python for Trust Collector, Policy Distributor, Policy Checker and ratings mechanism of our controllers. The Policy Checker and ratings mechanism were integrated in Ryu controller, whereas the Policy Distributor and Trust Collector were deployed as separate modules. Small number of controllers and OpenFlow switches were also deployed. We used the Emulab network evaluation testbed.

In the topology, we used one Policy Distributor, one Trust Collector, and varying number of controllers and switches were deployed for different evaluations. For the scalability tests, we used simulated switches and increased the number of controllers to up to 15, and the number of switches to up to 30. The configurations used are shown in Table 1 and the results are shown in Figure 3.

\begin{table}[h!]
\centering
\label{my-label}
\caption{Shows the different network configurations we created of controllers and switches.}
\begin{tabular}{|c|c|c|c|}
\hline
Configuration & Controllers & Switches & \begin{tabular}[c]{@{}c@{}}Malicious \\ controllers\end{tabular} \\ \hline
\hline
Config 1      & 5           & 10       & 2                                                                \\ \hline
Config 2      & 10          & 20       & 4                                                                \\ \hline
Config 3      & 15          & 30       & 6                                                                \\ \hline
\end{tabular}
\end{table}

For our correctness evaluations, we deliberately triggered one or more of the controllers to randomly install a malicious policy and then ran the rating mechanism in the controllers. All other controllers were able to detect the controller which installed wrong policy, and rated it negative. The Trust Collector aggregated the ratings from all these controllers. For all the tests we conducted, the scheme was always able to find the malicious controller(s) with zero false positives or false negatives.

Figure 3 shows the time taken to perform the entire process of rating and trust collection as the number of controllers involved in the process is increased. As seen from the graph, the time shows a linear pattern of increase and our scheme is able to work fast in finding out malicious controller.

\begin{figure}[h!]
  \includegraphics[width=\linewidth]{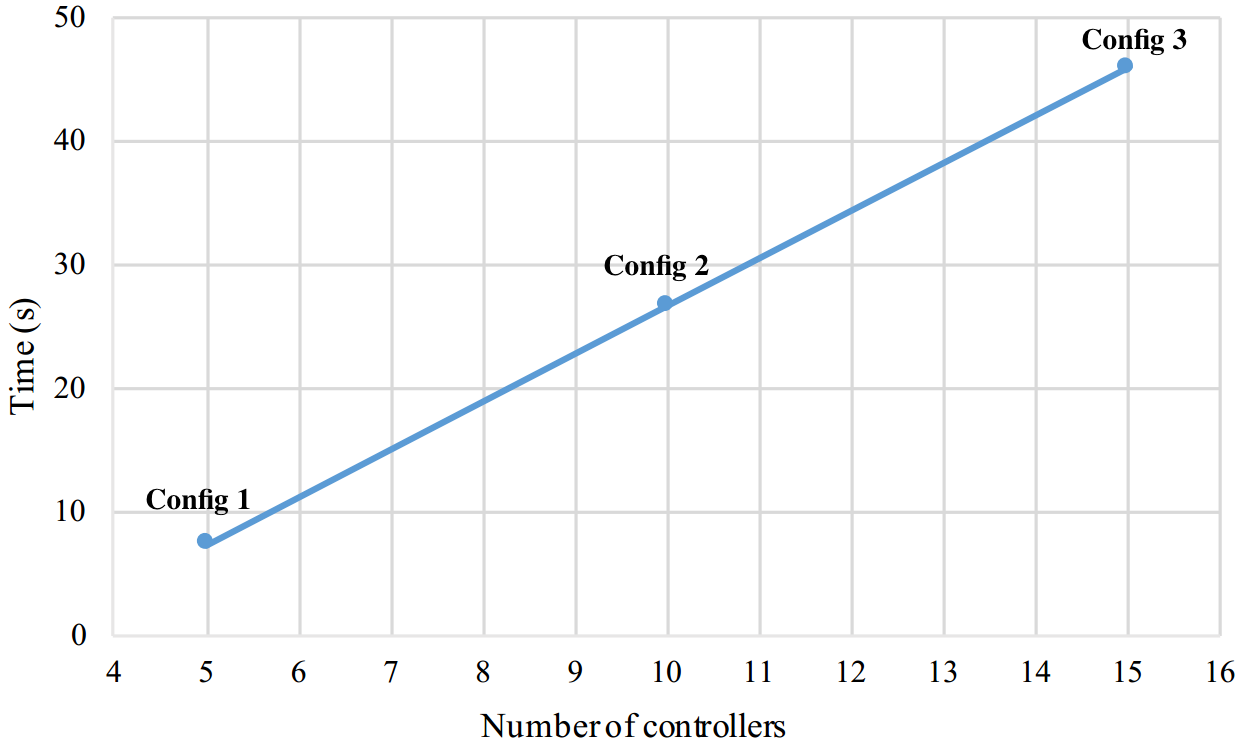}
  \caption{Scalability of the scheme: Shows the time taken in seconds for our entire rating and trust collection scheme to finish as the number of controller is increased.}
  \label{fig:boat1}
\end{figure}

The scheme defines a specific number of message exchanges (as shown in Figure 4) between the different components in the system, i.e. the controllers, switches, Trust Collector, and Policy Distributor. We use a centralized graph database, Neo4j [17], which serves as a 'noticeboard' for communication using the publish-subscribe mechanism. This saves network bandwidth since the Policy Distributor or Trust Collector do not have to broadcast messages containing commands such as 'startTrustCalculation', a command meant to be sent to all controllers to start the trust calculation process, to all controllers. Instead, the Trust Collector can publish this command by writing it in the centralized database and the controllers can read it from there. Thus only one message exchange has to be used instead of a broadcast of messages to all controllers. 

Neo4j has a Python library that handles the lower level network communication code and provides a RESTful web API which we invoke from our code to perform publish or subscribe functions. Note that each node in our evaluation setup has an IP address, this includes the node running Neo4j, and so the REST API can be invoked on the Neo4j database from any of the controllers and Trust Collector or Policy Ditributor components by using the IP of the Neo4j node.

The number of messages that need to be used for one complete process of our trust calculation is O(N*M) where N is the number of controllers involved in calculating the trust and M is the number of switches in the network, and there exists one instance of the Policy Distributor component and one instance of the Trust Collector component. While the Neo4j database based communication scheme described earlier helps get rid of broadcast messages, each controller (from N number of controllers) has to communicate with each of the switches (from the M number of total switches).

\begin{table}[h!]
\centering
\label{my-label}
\caption{Different network configurations we created of controllers and switches for scalability evaluation of number of messages. In each configuration, number of switches controlled by one controller is equal to (No. of Switches / No. of Controllers).}
\begin{tabular}{|c|c|c|c|}
\hline
Configuration & Controllers & Switches & \begin{tabular}[c]{@{}c@{}}Malicious \\ controllers\end{tabular} \\ \hline
\hline
Config 1 & 1 & 3 & 0 \\ \hline
Config 2 & 3 & 6 & 1 \\ \hline
Config 3 & 6 & 12 & 2 \\ \hline
Config 4 & 9 & 27 & 3 \\ \hline
\end{tabular}
\end{table}

Table 2 shows the various configurations of controllers and switches which we created in our evaluation setup. Note that this set of configurations created are different than those created for the earlier evaluation and were shown in Table 1. Figure 4 shows the number of messages that were used to perform one complete process of trust collections for the various configurations mentioned in Table 2. Note that here a message from a component A to component B is defined as one write of a message from a component A and its corresponding read by a component B. As can be seen from the graph, our scheme scales smoothly as the number of controllers and switches involved in the process is increased. For the highest configuration, Config 4, with 9 controllers and 24 switches, the scheme uses less than 250 messages to finish the entire process of trust calculation.

\begin{figure}[h!]
  \includegraphics[width=\linewidth]{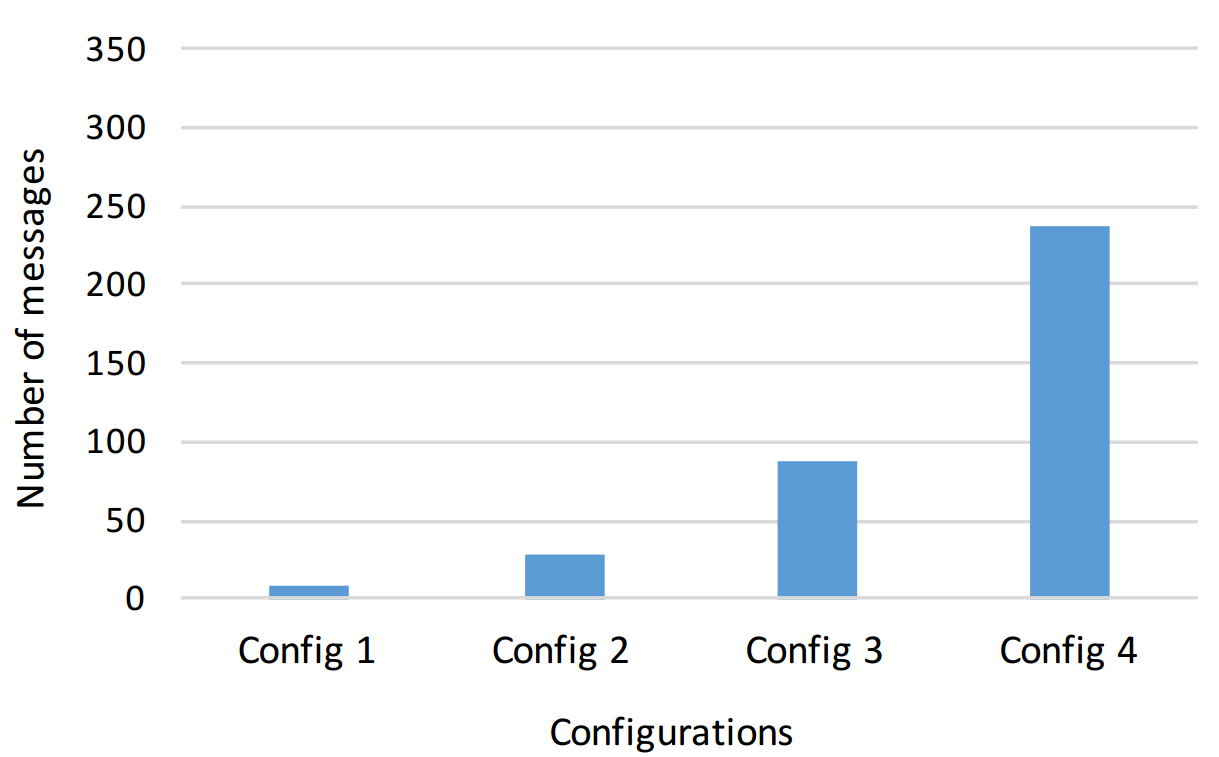}
  \caption{Shows the number of message exchanges that take place for different network configurations of controllers and switches, the names of the configurations on the X-axis (e.g. Config1, Config2, ..) refer to the configurations in Table 1.}
  \label{fig:boat1}
\end{figure}

\begin{table*}[]
\centering
\caption{Different network configurations we created of controllers and switches for bad mouthing evaluation. In each configuration, number of switches controlled by one controller is equal to (No. of Switches / No. of Controllers). And CX refers to the controller number, e.g. C1, C2, etc. The bad mouthing column shows the details of which controller(s) bad mouthed which other controller(s). The result column shows the final result after the trust calculation round which aggregates trust values from all controllers in the network. As see from results, the positive or negative majority ratings affect the result of whether a given controller is deemed trusted or untrusted.}
\label{my-label}
\begin{tabular}{|l|l|l|l|l|l|}
\hline
\multicolumn{1}{|c|}{Configuration} & \multicolumn{1}{c|}{Controllers} & Switches & \#Malicious controllers & Bad mouthing                 & Result                  \\ \hline
Config 1                                     & 3                                         & 6        & 1                       & C1 bad mouthed C2            & C2 found trusted        \\ \hline
Config 2                                     & 3                                         & 6        & 2                       & C1, C2 bad mouthed C3        & C3 found untrusted      \\ \hline
Config 3                                     & 6                                         & 12       & 1                       & C1 bad mouthed C2 and C3     & C2 and C3 found trusted \\ \hline
Config 4                                     & 6                                         & 12       & 2                       & C1, C2 bad mouthed C3 and C4 & C3 and C4 found trusted \\ \hline
\end{tabular}
\end{table*}

\section{Discussion}

Our test results presented earlier show that our scheme works correctly and efficiently in weeding out malicious controllers. Since the Trust Collector decides whether a controller is malicious based on aggregate of recommendations from all other controllers, therefore our scheme provides defense against bad mouthing attacks [18, 19]. 

In bad mouthing attacks, a malicious party provides dishonest recommendations for another good party to malign the name of the good party. But since our scheme does not make a decision of whether a controller is malicious based on recommendation from just one other controller, therefore we can provide defense against bad mouthing as long as malicious controllers are not the majority in the total number of deployed controllers. This assumption is reasonable since we can guarantee the number of controllers which would need to become malicious before the network collapses. That is, in a network with N controllers, our scheme is guaranteed to work correctly and identify malicious controllers as long as (N/2)+1 controllers stay uncompromised. This assumption is realistic since majority of controllers is unlikely to become malicious in an instant and if they become malicious one by one over time, then our scheme will identify the malicious controllers at all times when (N/2)+1 controllers are still uncompromised.

Our scheme of collecting ratings and aggregating trust and reputation using a Trust Collector component works more robustly and accurately than delegating trust and reputation management entirely to individual controllers in a distributed environment. This is because we always need a central entity that can aggregate the ratings generated by all the controllers that are part of the distributed environment, and the central entity can then make a decision of whether a given controller is malicious by looking at what the majority of ratings say about that controller. Alternatively, the central entity can also output the result of the ratings to a human operator who can decide whether a given controller is malicious based on both their domain knowledge about the network and also based on the majority of ratings that were received for that controller.

Introducing a central trust managing entity also helps in solving an important dilemma, which is, what happens if majority of rogue controllers vote against a controller which otherwise has installed correct policies? Let us examine a case of distributed trust management, in which controllers find the policy mismatches themselves, and there is no central entity for managing the trust and reputation. There are three controllers in a network, A, B, and C, each managing one switch under them, and connected to other switches too. A network administrator defines one flow rule, i.e. block any traffic originating from IP address 200.0.0.1. The controllers install the flow rules in their respective switches. After sometime, controllers probe the switches to find out whether other controllers installed correct policies in their respective switches. A finds out that B and C have (maliciously) installed flow rules in their switches which allow traffic originating from 200.0.0.1. It rates B and C negative. B and C on the other hand rate A as negative. In presence of an automated solution of shutting down or restricting a malicious controller, this will prove to be disastrous. If, however, a human operator has to approve the shutting down or restricting of a malicious controller, then it will be a burden for him to sort through the conflicting ratings of controllers against each other. 

Using the Trust Collector for aggregating the opinions about other controllers from each controller not only helps us in ascertaining the validity of recommendations with surety, but it also helps in eliminating broadcasts. If, for example, there are three controllers A, B, and C, then all of the nodes will have to send their reports to each other so that they can perform final trust calculation (since the final step in the trust calculation process inside a controller needs input from other controllers too). 

However, by introducing the Trust Collector in between, all controllers send their reports to this central entity, which simply forwards it to individual controllers. While it is true that our scheme introduces this one central point of compromise, the Trust Collector, but we emphasize that it is much easier to guard and protect one component if it can help us have a safe distributed environment of controllers where each controller does not have to guarded very well. As long as majority of controllers are not compromised, our scheme guarantees that the network will keep functioning correctly.

\section{Conclusion and Future Work}

Securing the controllers in SDN is an open problem for research. Researches carried so far do not address the problem of identifying malicious controllers, especially in multi-controller environment. We tackle this problem by employing a trust management scheme in which controllers rate each other on the basis correctness of flow rules installed by them in switches. We do this by introducing a centralized entity which keeps record of policies to be installed, so that controllers can compare them with installed policies for rating purpose. We coded the scheme and implemented it in Ryu controller as prototype and found that the scheme is able to successfully flag a malicious controller. Our next step will be to extensively evaluate our scheme in different network topologies and number of nodes, and also by launching various kinds of attacks on our trust management system.


\begin{thebibliography}{1}
\bibitem{IEEEhowto:kopka}
H. Kim, N. Feamster, Improving network management with software defined networking, Communications Magazine, IEEE 51 (2) (2013) 114-119.

\bibitem{IEEEhowto:kopka}
Hakiri, Akram, et al. "Software-defined networking: challenges and research opportunities for future internet." Computer Networks 75 (2014): 453-471.

\bibitem{IEEEhowto:kopka}
Sezer, Sakir, et al. "Are we ready for SDN? Implementation challenges for software-defined networks." Communications Magazine, IEEE 51.7 (2013): 36-43.

\bibitem{IEEEhowto:kopka}
Kreutz, Diego, et al. "Software-defined networking: A comprehensive survey." Proceedings of the IEEE 103.1 (2015): 14-76

\bibitem{IEEEhowto:kopka}
Kreutz, Diego, Fernando Ramos, and Paulo Verissimo. "Towards secure and dependable software-defined networks." Proceedings of the second ACM SIGCOMM workshop on Hot topics in software defined networking. ACM, 2013.

\bibitem{IEEEhowto:kopka}
Prasad, Abhinandan S., David Koll, and Xiaoming Fu. "On the Security of Software-Defined Networks." 2015 Fourth European Workshop on Software Defined Networks. IEEE, 2015.

\bibitem{IEEEhowto:kopka}
Liang, Zhengqiang, and Weisong Shi. "PET: A personalized trust model with reputation and risk evaluation for P2P resource sharing." Proceedings of the 38th annual Hawaii international conference on system sciences. IEEE, 2005.

\bibitem{IEEEhowto:kopka}
"Ryu SDN framework," https://osrg.github.io/ryu/

\bibitem{IEEEhowto:kopka}
"Emulab - Network Emulation Testbed," http://emulab.net/

\bibitem{IEEEhowto:kopka}
"Kaminsky DNS Attack," http://dankaminsky.com

\bibitem{IEEEhowto:kopka}
Shu, Zhaogang, et al. "Security in Software-Defined Networking: Threats and Countermeasures." Mobile Networks and Applications: 1-13.

\bibitem{IEEEhowto:kopka}
Porras, Phillip A., et al. "Securing the Software Defined Network  Control Layer." NDSS. 2015.

\bibitem{IEEEhowto:kopka}
Nunes, Bruno, et al. "A survey of software-defined networking: Past, present, and future of programmable networks." Communications Surveys and Tutorials, IEEE 16.3 (2014): 1617-1634.

\bibitem{IEEEhowto:kopka}
Phemius, Kévin, Mathieu Bouet, and Jérémie Leguay. "Disco: Distributed multi-domain sdn controllers." 2014 IEEE Network Operations and Management Symposium (NOMS). IEEE, 2014.

\bibitem{IEEEhowto:kopka}
Tootoonchian, Amin, and Yashar Ganjali. "HyperFlow: A distributed control plane for OpenFlow." Proceedings of the 2010 internet network management conference on Research on enterprise networking. 2010.

\bibitem{IEEEhowto:kopka}
Koponen, Teemu, et al. "Onix: A Distributed Control Platform for Large-scale Production Networks." OSDI. Vol. 10. 2010.

\bibitem{IEEEhowto:kopka}
"Neo4j Graph Database," http://neo4j.com

\bibitem{IEEEhowto:kopka}
S. Buchegger and J. L. Boudec. "Coping with False Accusations in Misbehavior Reputation Systems for Mobile Ad-Hoc Networks."" EPFL tech. rep. IC/2003/31, EPFL-DI-ICA, 2003.

\bibitem{IEEEhowto:kopka}
Sun, Yan, Zhu Han, and KJ Ray Liu. "Defense of trust management vulnerabilities in distributed networks." IEEE Communications Magazine 46.2 (2008): 112-119.

\end{thebibliography}
\end{document}